\documentclass[prl,nofootinbib,superscriptaddress,twocolumn]{revtex4}
\def\mysection#1{{\bf #1.} }

\usepackage{amssymb}
\usepackage{amsmath}
\usepackage{graphicx}
\usepackage{longtable}
\usepackage{verbatim}
\usepackage{amsfonts}
\usepackage{hyperref}

\newcommand{\arXiv}[2]{\href{http://arxiv.org/pdf/#1}{{\tt [#2/#1]}}}
\newcommand{\arXivold}[1]{\href{http://arxiv.org/pdf/#1}{{\tt [#1]}}}

\newcommand{\be}{\begin{equation}}
\newcommand{\ee}{\end{equation}}
\newcommand{\bea}{\begin{eqnarray}}
\newcommand{\eea}{\end{eqnarray}}
\newcommand{\beq}{\begin{equation}}
\newcommand{\eeq}{\end{equation}}
\def\beqa{\begin{eqnarray}}

\def\eeqa{\end{eqnarray}}

\def\lsim{\mathrel{\rlap{\lower4pt\hbox{\hskip0.5pt$\sim$}}
    \raise1pt\hbox{$<$}}}         
\def\gsim{\mathrel{\rlap{\lower4pt\hbox{\hskip0.5pt$\sim$}}
    \raise1pt\hbox{$>$}}}         

\begin{document}

\vspace*{-30mm}

\title{A note on Inflation and the Swampland }
\vskip 0.3cm
\author{A. Kehagias }
\address{Physics Division, National Technical University of Athens, \\15780 Zografou Campus, Athens, Greece}
\author{A. Riotto}
\address{D\'epartement de Physique Th\'eorique and Centre for
  Astroparticle Physics (CAP), Universit\'e de Gen\`eve, 24 quai E. Ansermet, CH-1211 Geneva, Switzerland
}

\date{\today}

%
%

\begin{abstract}
\noindent
 We provide some comments about the  constraints  on the inflationary  models inferred from the two  Swampland criteria which have been recently proposed.
 In particular we argue that,  in the absence of any knowledge about the origin of the adiabatic curvature perturbations, within the slow-roll single field models of inflation there is no tension  between the swampland criteria and  the current lower bound on the tensor-to-scalar ratio.

\end{abstract}

\maketitle

\noindent
\mysection{\it Introduction}
\noindent
String theory provides an immense landscape of vacua. Such landscape,  although populated by consistent low-energy effective theories,
is  surrounded by the Swampland \cite{vafa-3,vafa-2,vafa-1,vafa0,vafa1}, a region where inconsistent semiclassical effective theories live. 
In other words, the string landscape represents a small portion of seemingly consistent effective theories of the Swampland where  additional conditions are satisfied. Such conditions are provided for example the weak gravity conjecture \cite{vafa-3,arkani}. 

A particular issue concerns the maximally symmetric vacua and in particular the Minkowski, the anti-de Sitter (AdS) and the de Sitter (dS) spacetimes. As  it looks  difficult to obtain dS vacua contrary to the vast of Minkowski and AdS backgrounds in string theory, it has been proposed that dS is not part of the landscape, but rather lives in the Swampland \cite{carta,dd,vafa-2}.  Indeed, although supersymmetric (or non-supersymmetric)  Minkowski and AdS solutions in string theory are common, which are also standard vacuum solutions of the low-energy effective supergravity theory, dS ones are rare and difficult to construct. It is then natural to suspect that dS space does not reside in the landscape.
 
 A criterion that make the  dS vacua to  be part of the  Swampland has been proposed in Ref. \cite{vafa-2} and  is one of the two Swampland criteria  proposed in order to  specify if an effective theory is part of the Swampland or if it shares properties of a consistent quantum gravity theory.  
These two criteria are as follows:
\begin{itemize}
\item
 {\it C1}: Scalar field excursions  in  reduced Planck units $M_p=(8\pi G)^{-1/2}\simeq 2.4\times 10^{18}$ GeV  in field space are bounded from above  \cite{vafa-2}
\begin{eqnarray}
 \frac{\Delta \phi}{M_p}< {\cal O}(1).  \label{c1}
 \end{eqnarray} 
 
 \item 
{\it  C2}: The gradient of the potential in any direction of a canonically normalised scalar field in a consistent gravity theory satisfies the bound  \cite{vafa1}
\begin{eqnarray}
M_p\frac{|\nabla_\phi V|}{V}>c\sim {\cal O}(1), \label{c2}
\end{eqnarray}
 whenever $V>0$. 
 \end{itemize}
 Of course, as it has been noticed already in Ref. \cite{dvali}, as long as the parameter  $c$ is not specified,
 the criterion (\ref{c2})   is contentless. However, for the sake of the argument, we are going to assume its validity  in order to discuss its cosmological implications.
Some of them have been discussed in Ref. \cite{vafa1,Blum0,Heb,Cicoli,Lust,Grimm,Heid,Blum,Taylor,Blum2,Shiu,andriot,today} and these implications   for  the paradigm of  inflation \cite{guth81,lrreview}  will be the subject of this short note.
\vskip0.5cm
\noindent
\mysection{\it Inflation and the Swampland}
\noindent Inflation is currently  the dominant paradigm 
providing  the  
initial seeds for the Cosmic Microwave Background (CMB) anisotropies and the Large-Scale-Structure (LSS).

 In slow-roll single field models of  inflation, the    primordial density
 fluctuations are generated when quantum fluctuations leave the Hubble radius and are left imprinted on super-Hubble scales \cite{muk81,bardeen83}. 

 The power spectrum of the gauge-invariant curvature perturbations reads

\begin{eqnarray}
{\cal P}_\zeta(k)=\frac{1}{2M_p^2 \epsilon}\left(\frac{H}{2\pi}\right)^2\left(\frac{k}{aH}\right)^{n_\zeta-1}
\end{eqnarray}
where $n_\zeta\approx 1$ is the spectral index,  

\be
\epsilon=\frac{\dot{\phi}^2}{2H^2M_p^2}
\ee
 is one of the slow-roll parameters and $H=\dot a/a$
is the Hubble parameter. 
 Similarly,
the 
power spectrum of tensor modes turns out to be

\begin{equation}
{\cal P}_{T}(k)=\frac{8}{M_p^2}\left(\frac{H}{2\pi}\right)^2
\left(\frac{k}{aH}\right)^{-2\epsilon}.
\end{equation}
The power spectra are roughly constant on relevant CMB anisotropy scales 
and we  may define a tensor-to-scalar amplitude ratio 

\begin{equation}
\label{q}
r=\frac{{\cal P}_{T}(k)}{{\cal P}_{\zeta}(k)}=16\epsilon\, .
\end{equation}
Furthermore, assuming slow-roll, the parameter $\epsilon$ reduces to

\be
\epsilon=\frac{M_p^2}{2}\left(\frac{V'}{V}\right)^2,
\ee
where $V(\phi)$ is the inflaton potential and the prime stands for differentiation with respect to the inflaton field.
The  inflaton field excursion can be expressed in terms of the slow-parameter $\epsilon$

\begin{equation}
\label{k}
\frac{1}{M_p}\left|\frac{{\rm d}\phi}{{\rm d}N}\right|=\sqrt{2\epsilon},
\end{equation}
where ${\rm d}\phi$ is the change in the inflaton field in ${\rm d}N=H{\rm d}t\simeq
{\rm d}\,{\rm ln}\, a$ Hubble times. 

The link between the slow-roll parameter $\epsilon$ and the number of e-folds $N$ till the end of inflation is model dependent, but one can roughly parametrises it as \cite{lrreview}

\be
\epsilon\simeq \frac{1}{N^p},
\ee
from which one deduces a field excursion of the form \cite{vafa1}

\be
\frac{\Delta \phi}{M_p}\simeq \sqrt{2} N^{1-p/2}.
\ee
From this expression one can see that, if the  criterion {\it C1} is taken seriously, one might have troubles in getting a large enough number of e-folds for the inflating patch to encompass our current horizon (that is $N\simeq 60$). However, a large number of e-folds is easily obtained in models with a plateau. This point was already noticed in Ref. \cite{vafa1}. Take for example a potential of the form

\be
\label{pot}
V(\phi)=V_0\left(1-e^{-q \phi/M_p}\right),
\ee
where $q$ is positive. Supposing that during inflation  $V_0$ dominates,  the parameter $\epsilon$ is

\be
\epsilon=\frac{1}{2}\frac{1}{q^2 N^2}
\ee
and the excursion of the inflaton field reads

\be
\frac{\Delta \phi}{M_p}\simeq \frac{1}{q}\ln N.
\ee
Essentially, in the inflationary models characterized by a potential of the form   (\ref{pot}), one can  easily obtain a large number of e-folds since  $\epsilon$ is extremely tiny and 
excursions in the inflaton field can be easily sub-Planckian  for  $q\gsim \ln 60\simeq 4$.
 
Models with an exponentially suppressed plateau may  be naturally obtained 
even if the  original
potential is not particularly flat, but the kinetic term of the inflaton is not canonical and  becomes singular
at some  field value. This was  originally  observed in Ref.   \cite{stewart} and further discussed in Ref. \cite{lrreview}. Recently this class of attractor models  has been thoroughly investigated and extended \cite{at}. They  have the peculiarity that at large values of the inflaton field any potential $V(\varphi)$, upon  converting the non-canonically normalized field $\varphi$ into to a canonically normalized
inflaton field $\phi$,  acquires a plateau at large values of $\phi$. Take for example an inflationary model whose  Lagrangian reads \cite{stewart}

\be
{\cal L}=\frac{1}{2}\frac{(\partial\varphi)^2}{\left(1-q^2\varphi^2/4M_p^2\right)^2}-\frac{1}{2}\mu^2\varphi^2, 
\label{non-min}
\ee
where $\mu$ is a mass which can be even identified with the cut-off scale $M$.
The canonically normalized field is 

\be
\label{cn}
\varphi=\frac{2M_p}{q}\,{\rm tanh}\frac{q\,\phi}{2M_p}
\ee
and the corresponding Lagrangian becomes

\be
{\cal L}=\frac{1}{2}(\partial\phi)^2-\frac{1}{2}V_0\, {\rm tanh}^2\frac{q\,\phi}{2M_p}\,\,\,,\,\,\,\,V_0=\frac{2}{q^2}\mu^2M_p^2.
\ee
At large values of the $\phi$ field the inflationary model is simply  described by a dominating vacuum energy plateau plus an exponential suppressed field contribution. We conclude that the Swampland criterion 
{\it C1} can be easily satisfied  for a large class of inflationary models and therefore it does not represent a threat to the inflationary paradigm.

What about the criterion {\it C2}? It implies

\be
\label{ddd}
\epsilon>\frac{c^2}{2}.
\ee
This lower bound, if taken at face value, would rule out all slow-roll single field models of inflation as an accelerated period can take place only if 
$\epsilon<1$. However, 
the requirement that no critical dS vacua exist, even meta-stable ones, as proposed in Ref. \cite{vafa-2},  allows any value of $c$ (as long as $c>0$), even tiny values.
Given  that no quantitive argument for having $c\simeq 1$ is currently available,  $c\simeq 10^{-1}$ seems  as good as $c\simeq 1$ to us\footnote{A bound of the form $M_p|\nabla V|>V^2/M_p^4$ as discussed in Ref. \cite{dvali} would lead  
$c\sim  (H/M_p)^2\ll 1$, making  all the discussion irrelevant.}. If so,   $\epsilon$ needs to be  larger than $5\cdot 10^{-3}$ and this value is small enough to give rise to a period of inflation (and still be in agreement with having the spectral index $n_\zeta\simeq 1$).

The real problem with the lower bound (\ref{ddd}) is that in slow-roll single field  models of inflation the consistency relation (\ref{q}) holds and therefore the current bound $r\lsim 0.07$ (from the absence of the tensor mode-induced B-mode polarisation in the CMB anisotropies \cite{pl}) converts into the upper bound

\be
\epsilon\lsim 4.4\cdot 10^{-3},
\ee
which is in clash with the criterion {\it C2} even allowing $c={\cal  O}(10^{-1})$.  Let us elaborate about this particular argument  in the next section.
\vskip 0.5cm
\noindent
\mysection{\it Avoiding the Swampland}
\noindent
The authors of Ref. \cite{vafa1} concluded that  inflationary models are generically in tension with   the Swampland criterion {\it C2}. However, we do not share the same opinion. What do we really know about the  perturbations generated during inflation? We believe it is fair to say that: 

\begin{itemize}

\item scalar and tensor  perturbations are almost  scale-invariant
and scalar perturbations are  adiabatic;

\item scalar perturbations are nearly-Gaussian, the level of non-Gaussianity being 
 severely constrained \cite{ng}.

\end{itemize}
These are  observational facts which may not be disputed. 
The somewhat gloomy  reality that we have to accept   is that  we do not know
 what is the real  source of the scalar perturbations during inflation. 
 Indeed, 
even though the  inflationary paradigm is quite simple to spell, the source of the 
 cosmological adiabatic perturbations  remains a mystery  (and probably it will remain  unless a large level of non-Gaussianity is observed).
 
It is 	imaginable that the adiabatic scalar fluctuations are not due to the inflaton fluctuations (in the flat gauge) and therefore 
the curvature perturbation $\zeta$  does not remain  constant on super-Hubble scales. 
 Quite the opposite,  $\zeta$ can vary   due  to a non-adiabatic
pressure perturbation density $\delta P_{\rm nad}$  which may be 
appear if  extra  degrees of freedom are present. The corresponding evolution equation 
is

\be
\dot\zeta=\frac{H}{\overline \rho+\overline P}\,\delta P_{\rm nad},
\ee
where $\overline \rho$ and $\overline P$ are the background energy and pressure densities, respectively.
If so, the final adiabatic cosmological perturbation $\zeta$ finds its origin not from a single-clock degree of freedom and the consistency relation (\ref{q}) is not valid any longer. 

The curvaton \cite{curvaton1,LW,curvaton3,LUW} provides an example of a mechanism in which the curvature perturbation is generated not at Hubble crossing, but on super-Hubble scales when the
curvaton isocurvature perturbations are converted into curvature perturbations upon the curvaton decay (after the end of inflation when radiation is present). In such a scenario the non-adiabatic
pressure perturbation density $\delta P_{\rm nad}$ is given by

\be
\delta P_{\rm nad}=\frac{4\overline{\rho}_{\rm rad}\overline{\rho}_{\sigma}}{4\overline{\rho}_{\sigma}+3\overline{\rho}_{\rm rad}}\zeta_\sigma,
\ee
where $\sigma(\vec x,t)$ is the nearly massless curvaton field and 

\be
\zeta_\sigma=H\frac{\delta\rho_\sigma}{\dot{\overline\rho}_\sigma}
\ee
is the curvature curvaton perturbation in the flat gauge.

The curvaton mechanism is not the only option. For instance,  the modulated decay scenario \cite{gamma1,gamma2} takes advantage of the fact that 
 one can generate   spatial
fluctuations  in the decay rate of the inflaton field on  super-Hubble scales,  if the  decay rate has a dependence on the vacuum expectation value of a nearly massless scalar field.
The fluctuations of the latter induce a perturbation in the inflaton decay rate and, consequently, in the temperature of the 
radiation generated at decay.  

Last, but not least,  it may happen that the curvature perturbation  received the dominant contribution at the last phase of inflation \cite{end1,end2}.
This happens if  inflation ends when the inflaton field  acquires some critical value $\phi_e$ and such value depends on some other scalar field $\sigma$. If the latter is again nearly massless
during inflation, the curvature perturbation will read

\be
\zeta\simeq \frac{H}{\dot\phi_e}\frac{\partial \phi_e}{\partial\sigma}\delta\sigma.
\ee
In all these scenarios the consistency relation (\ref{q}) is not valid and therefore one may not connect the tensor-to-scalar ratio $r$ to the slow-roll parameter $\epsilon$. This implies that
the criterion {\it C2}, does not provide any information. In fact, in these mechanisms to generate the primordial curvature perturbation $\zeta$, the contribution to it from the
inflaton field is suppressed by taking a low value of the Hubble parameter during inflation. One therefore does not expect a measurable amount of tensor modes through the B-mode polarisation of the CMB anisotropies. 

\vskip 0.5cm
\noindent
\mysection{\it Conclusions} In this short note we have made the point that the Swampland criteria are not in tension with the inflationary paradigm as long as we do not have the certainty that the parameter $c$ is unity and that the curvature perturbation is originated by a single-clock. A future  detection of a large level of non-Gaussianity, which is incompatible with slow-roll single field models of inflation,
 will put our reasoning on firmer grounds. In fact, if taken seriously, the Swampland criteria might suggest that the curvaton-like mechanisms to give origin to the curvature perturbation are to be preferred.
 On the other hand, a measurement of a high level of tensor modes will make our argument less defendable.
 \vskip 0.5cm
  \noindent
{\it Acknowledgment} A.R. is supported by the Swiss National Science Foundation (SNSF), project Investigating {\it the Na-
ture of Dark Matter}, project number: 200020-159223. A.K. is supported by the GSRT under the EDEIL/67108600 of NTUA. 



\begin{references}

\bibitem{vafa-3} 
  C.~Vafa,
  ``The String landscape and the Swampland,''
  hep-th/0509212.

\bibitem{vafa-2} 
  H.~Ooguri and C.~Vafa,
  ``Non-supersymmetric AdS and the Swampland,''
  Adv.\ Theor.\ Math.\ Phys.\  {\bf 21}, 1787 (2017)
  \arXiv{1610.01533}{hep-th}.

\bibitem{vafa-1}
  T.~D.~Brennan, F.~Carta and C.~Vafa,
  ``The String Landscape, the Swampland, and the Missing Corner,''
  \arXiv{1711.00864}{hep-th}.

\bibitem{vafa0} 
  G.~Obied, H.~Ooguri, L.~Spodyneiko and C.~Vafa,
  ``De Sitter Space and the Swampland,''
  \arXiv{1806.08362}{hep-th}.
\bibitem{vafa1} 
  P.~Agrawal, G.~Obied, P.~J.~Steinhardt and C.~Vafa,
  ``On the Cosmological Implications of the String Swampland,''
  \arXiv{1806.09718}{hep-th}.


\bibitem{arkani}
  N.~Arkani-Hamed, L.~Motl, A.~Nicolis and C.~Vafa,
  ``The String landscape, black holes and gravity as the weakest force,''
  JHEP {\bf 0706} (2007) 060
  \arXivold{hep-th/0601001}.

\bibitem{carta} 
  T.~D.~Brennan, F.~Carta and C.~Vafa,
  ``The String Landscape, the Swampland, and the Missing Corner,''
  \arXiv{1711.00864}{hep-th}.

\bibitem{dd} 
  U.~H.~Danielsson and T.~Van Riet,
  ``What if string theory has no de Sitter vacua?,''
  \arXiv{1804.01120}{hep-th}.

\bibitem{dvali} 
  G.~Dvali and C.~Gomez,
  ``On Exclusion of Positive Cosmological Constant,''
  \arXiv{1806.10877}{hep-th}.

\bibitem{Blum0} 
  R.~Blumenhagen, I.~Valenzuela and F.~Wolf,
  ``The Swampland Conjecture and F-term Axion Monodromy Inflation,''
  JHEP {\bf 1707}, 145 (2017)
  \arXiv{1703.05776}{hep-th}.

\bibitem{Heb} 
  A.~Hebecker, P.~Henkenjohann and L.~T.~Witkowski,
  ``Flat Monodromies and a Moduli Space Size Conjecture,''
  JHEP {\bf 1712}, 033 (2017)
  \arXiv{1708.06761}{hep-th}.

\bibitem{Lust} 
  D.~Lust and E.~Palti,
  ``Scalar Fields, Hierarchical UV/IR Mixing and The Weak Gravity Conjecture,''
  JHEP {\bf 1802}, 040 (2018)
  \arXiv{1709.01790}{hep-th}.

\bibitem{Cicoli} 
  M.~Cicoli, D.~Ciupke, C.~Mayrhofer and P.~Shukla,
  ``A Geometrical Upper Bound on the Inflaton Range,''
  JHEP {\bf 1805}, 001 (2018)
  \arXiv{1801.05434}{hep-th}.

\bibitem{Grimm} 
  T.~W.~Grimm, E.~Palti and I.~Valenzuela,
  ``Infinite Distances in Field Space and Massless Towers of States,''
  \arXiv{1802.08264}{hep-th}.

\bibitem{Heid} 
  B.~Heidenreich, M.~Reece and T.~Rudelius,
  ``Emergence and the Swampland Conjectures,''
  \arXiv{1802.08698}{hep-th}.  
  

  
\bibitem{Blum} 
  R.~Blumenhagen,
  ``Large Field Inflation/Quintessence and the Refined Swampland Distance Conjecture,''
  \arXiv{1804.10504}{hep-th}.
  
  
\bibitem{Taylor} 
  W.~Taylor and A.~P.~Turner,
  ``An infinite swampland of U(1) charge spectra in 6D supergravity theories,''
  JHEP {\bf 1806}, 010 (2018)
  \arXiv{1803.04447}{hep-th}.  
  
\bibitem{Blum2} 
  R.~Blumenhagen, D.~Kläwer, L.~Schlechter and F.~Wolf,
  ``The Refined Swampland Distance Conjecture in Calabi-Yau Moduli Spaces,''
  JHEP {\bf 1806}, 052 (2018)
  \arXiv{1803.04989}{hep-th}. 
 

\bibitem{Shiu} 
  A.~Landete and G.~Shiu,
  ``Mass Hierarchies and Dynamical Field Range,''
  \arXiv{1806.01874}{hep-th}.
  
  \bibitem{andriot} 
  D.~Andriot,
  ``On the de Sitter Swampland criterion,''
  \arXiv{1806.10999}{hep-th}.


\bibitem{today}
  A.~Achúcarro and G.~A.~Palma,
  ``The string swampland constraints require multi-field inflation,''
  arXiv:1807.04390 [hep-th].

\bibitem{guth81} A. Guth, 
``The Inflationary Universe: A Possible Solution to the Horizon and Flatness Problems,''
Phys. Rev. D {\bf 23}, 347 (1981)
\bibitem{lrreview} D.~H.~Lyth and A.~Riotto,
  ``Particle physics models of inflation and the cosmological density perturbation,''
  Phys.\ Rept.\  {\bf 314}, 1 (1999)
  \arXivold{hep-ph/9807278};\\
  A.~Riotto,
  ``Inflation and the theory of cosmological perturbations,''
  ICTP Lect.\ Notes Ser.\  {\bf 14}, 317 (2003)
  \arXivold{hep-ph/0210162};\\
   W.~H.~Kinney,
  ``Cosmology, inflation, and the physics of nothing,''
  NATO Sci.\ Ser.\ II {\bf 123}, 189 (2003)
  \arXivold{astro-ph/0301448}.
\bibitem{muk81} V.~F.~Mukhanov and G.~V.~Chibisov,
  ``Quantum Fluctuations and a Nonsingular Universe,''
  JETP Lett.\  {\bf 33}, 532 (1981)
  [Pisma Zh.\ Eksp.\ Teor.\ Fiz.\  {\bf 33}, 549 (1981)].
\bibitem{bardeen83} J. M. Bardeen, P. J. Steinhardt, and M. S. Turner, 
``Spontaneous Creation of Almost Scale - Free Density Perturbations in an Inflationary Universe,''
  Phys.\ Rev.\ D {\bf 28}, 679 (1983).





\bibitem{stewart}  E.~D.~Stewart,
  ``Inflation, supergravity and superstrings,''
  Phys.\ Rev.\ D {\bf 51}, 6847 (1995)
 \arXivold{hep-ph/9405389}.

\bibitem{at}  R.~Kallosh, A.~Linde and D.~Roest,
  ``Superconformal Inflationary $\alpha$-Attractors,''
  JHEP {\bf 1311}, 198 (2013)
  \arXiv{1311.0472}{hep-th}.



\bibitem{pl} P.~A.~R.~Ade {\it et al.} [BICEP2 and Keck Array Collaborations],
  ``Improved Constraints on Cosmology and Foregrounds from BICEP2 and Keck Array Cosmic Microwave Background Data with Inclusion of 95 GHz Band,''
  Phys.\ Rev.\ Lett.\  {\bf 116}, 031302 (2016)
  \arXiv{1510.09217}{astro-ph.CO}.



\bibitem{ng} P.~A.~R.~Ade {\it et al.} [Planck Collaboration],
  ``Planck 2015 results. XVII. Constraints on primordial non-Gaussianity,''
  Astron.\ Astrophys.\  {\bf 594}, A17 (2016)
  \arXiv{1502.01592}{astro-ph.CO}.

\bibitem{curvaton1} K.~Enqvist and M.~S.~Sloth,
``Adiabatic CMB perturbations in pre big bang string cosmology,''
Nucl.\ Phys.\ B {\bf 626}, 395 (2002)
\arXivold{hep-ph/0109214}.




\bibitem{LW}
D.~H.~Lyth and D.~Wands,
``Generating the curvature perturbation without an inflaton,''
Phys.\ Lett.\ B {\bf 524}, 5 (2002)
\arXivold{hep-ph/0110002}.

\bibitem{curvaton3} T.~Moroi and T.~Takahashi,
``Effects of cosmological moduli fields on cosmic microwave background,''
Phys.\ Lett.\ B {\bf 522}, 215 (2001)
[Erratum-ibid.\ B {\bf 539}, 303 (2002)]
\arXivold{hep-ph/0110096}.






\bibitem{LUW}
D.~H.~Lyth, C.~Ungarelli and D.~Wands,
  ``The Primordial density perturbation in the curvaton scenario,''
  Phys.\ Rev.\ D {\bf 67}, 023503 (2003)
  \arXivold{astro-ph/0208055}.

\bibitem{gamma1} G.~Dvali, A.~Gruzinov and M.~Zaldarriaga,
  ``A new mechanism for generating density perturbations from inflation,''
  Phys.\ Rev.\ D {\bf 69}, 023505 (2004)
\arXivold{astro-ph/0303591}.

\bibitem{gamma2} 
L.~Kofman,
\arXivold{astro-ph/0303614}.


  
  \bibitem{end1}
 D.~H.~Lyth,
  ``Generating the curvature perturbation at the end of inflation,''
  JCAP {\bf 0511}, 006 (2005)
  \arXivold{astro-ph/0510443}.
    
  \bibitem{end2} D.~H.~Lyth and A.~Riotto,
  ``Generating the Curvature Perturbation at the End of Inflation in String Theory,''
  Phys.\ Rev.\ Lett.\  {\bf 97}, 121301 (2006)
  \arXivold{astro-ph/0607326}.



















































\end{references}
\end{document}